\documentclass[journal]{IEEE}
\jmonth{May}
\pubyear{2022}

\usepackage{graphics}
\usepackage{listings}
\usepackage{xcolor}
\usepackage{float}
\usepackage[font=bf]{caption}
\graphicspath{ {images/} }
\usepackage{subcaption}

\definecolor{codegreen}{rgb}{0,0.6,0}
\definecolor{codegray}{rgb}{0.5,0.5,0.5}
\definecolor{codepurple}{rgb}{0.2,0,0.95}
\definecolor{backcolour}{rgb}{0.95,0.95,0.95}

\lstdefinestyle{mystyle}{
    backgroundcolor=\color{backcolour},   
    commentstyle=\color{codegreen},
    keywordstyle=\color{codepurple},
    numberstyle=\tiny\color{codegray},
    stringstyle=\color{codepurple},
    basicstyle=\ttfamily\footnotesize,
    breakatwhitespace=false,         
    breaklines=true,                 
    captionpos=b,                    
    keepspaces=true,                 
    numbers=left,                    
    numbersep=5pt,                  
    showspaces=false,                
    showstringspaces=false,
    showtabs=false,                  
    tabsize=2
}

\begin{document}

\title{FPGA Random Number Generator}
\author{Jacob Hammond}
\affil{Johns Hopkins University}  
\maketitle
\begin{center}
May 4th, 2022
\end{center}

\begin{abstract}
Random number generation is a key technology that is useful in a variety of ways. Random numbers are often used to generate keys for data encryption. Random numbers generated at a sufficiently long length can encrypt sensitive data and make it difficult for another computer or person to decrypt the data. Other uses for random numbers include statistical sampling, search/sort algorithms, gaming, and gambling. Due to the wide array of applications for random numbers, it would be useful to create a method of generating random numbers reliably directly in hardware to generate a ready supply of a random number for whatever the end application may be. This paper offers a proof-of-concept for creating a verilog-based hardware design that utilizes random measurement and scrambling algorithms to generate 32-bit random synchronously with a single clock cycle on a field-programmable-gate-array(FPGA). 
\end{abstract}

\pagebreak
\section{Introduction}

Due to the extensive applications of random number generators, there is a need for processors to be able to provide this capability in a wide variety of applications. It would be useful to create a common logical design that could be dropped into existing computers to supply random numbers quickly and efficiently. The goal of this project aims to create a hardware implementation of a random number generator on a FPGA and demonstrate the randomness of the generated numbers. 

\subsection{Relevance to computer Architecture}
The key topics that are covered in this project that are related to computer architecture are as followed. 

\subsubsection{Algorithm Implementation on hardware}
One of the key skills in computer architecture is translating a mathematical algorithm into concrete hardware functions. This is done by using the many computer arithmetic functions such as multiplication, division, addition, shifts, gates, counters, AND/OR/XOR/NOT and other logical gates. These functions will have to be used combinationally to embody a mathematical algorithm to generate random numbers via hardware. 

\subsubsection{Utilizing I/O inputs on an FPGA}
In the case of generating true-random numbers, often an analog measurement must take place. This is a great opportunity to use the Nexys A7 development board and interface with the I/O to obtain such a measurement and route that input into the design and compute with it. I/O is the backbone of how every computer communicates with its end-user, so demonstrating communication with I/O devices to obtain inputs and compute an output is a useful skill that can be applied to many computing applications. 

\subsubsection{Finite State Machines}
By using combinational logic, this project will result in a finite state machine that performs the predetermined action of generating random numbers. By observing the state over each iteration, data outputs can be analyzed and displayed in creative ways. 

\subsubsection{Linear feedback in computer design}
Many random number generators utilize algorithms with some sort of linear feedback. Linear feedback is unique in that the input is a linear function of the previous state. This property in computer design will allow us the ability to create random number generators that produce a continuous stream of outputs synchronously with the clock. 

\section{Background}
Random number generation is a key technology that is useful in a variety of ways. Random numbers are often used to generate keys for data encryption. Random numbers at a sufficiently long length can encrypt sensitive data and make it difficult for another computer or person to decrypt the data. Other uses for random numbers are statistical sampling, computer simulation, and search/sort algorithms, gaming, and even casino/gambling machines.

\subsection{Types of Random Numbers}
There is an important distinction when it comes to generating random numbers---random number generators fall under two sub-types: pseudo-random, or true-random. 

\subsubsection{Pseudo-random numbers}
Pseudo-random numbers are generated via a mathematical algorithm and some starting value known as a “seed” to create a finite number of random numbers. The limit of these algorithms can be large, but after so many iterations repeating values are produced. Generating pseudo-random numbers happens entirely within a computer and can be generated in hardware/software without needing any external inputs.\cite{lecuyer}

\begin{figure}[H]
\centering
\includegraphics[scale=0.30]{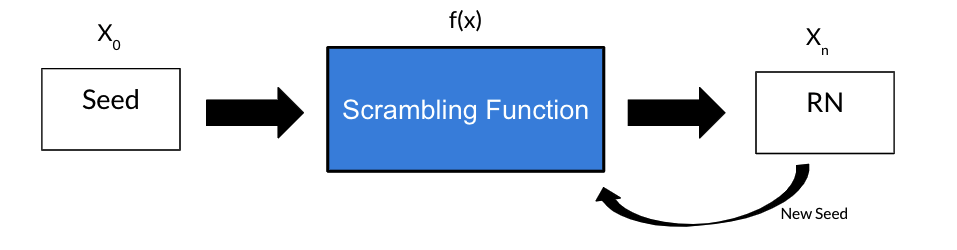}
\caption{High-level Block Diagram of a Pseudo-Random Number Generator}
\label{fig:pseudo_block}
\end{figure}

\subsubsection{True-random numbers}
True-random numbers are considered truly random and are generated using a physical entity of randomness. This might be something like radioactive decay/entropy or electrical noise which are physically random events in the real world. Computers can take readings of these events via sampling, and produce a random number utilizing an algorithm to produce a random number.\cite{lecuyer}

\begin{figure}[H]
\centering
\includegraphics[scale=0.30]{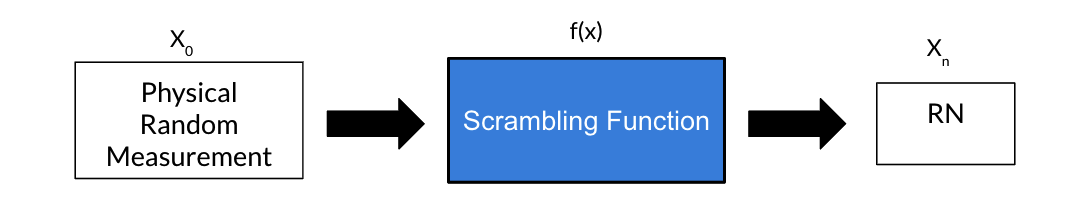}
\caption{High-level Block Diagram of a True-Random Number Generator}
\label{fig:true_block}
\end{figure}

\section{Technical Approach}
The first step of demonstrating a solution to this challenge is to outline a system architecture block diagram that defines the function and flow of the individual components of the FPGA random number generator.

\begin{figure}[H]
\centering
\includegraphics[scale=0.30]{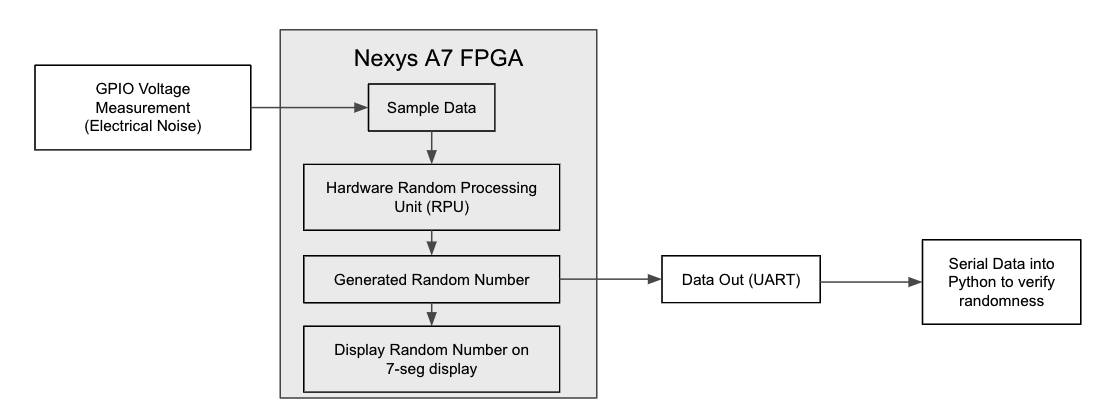}
\caption{Block Diagram - Nexys A7 Random Number Generator}
\label{fig:scope_block}
\end{figure}

This design has several foundational blocks. First a voltage measurement must be taken utilizing the GPIO inputs on the Nexys A7 board. This will be acheived by sampling an analog voltage signal into a digital number via an analog-to-digital-converter (ADC). Then that voltage measurement will be fed into the random processing unit as a seed --- which will be the scrambling functions that generates the random number of desired length. Next the random number is passed to two places: the 7 segment display of the Nexys A7 board, and to a UART transmit module which will send the random number to a connected computer. The connected computer will then use python to evaluate the numbers generated for statistical randomness. 

\subsection{Description of Work}
The process to create the working design can be divided into 5 distinct functional modules. 
\begin{enumerate}
    \item Verilog module that generates random numbers with specific algorithms.
    \item Verilog module that reads the Nexys A7 GPIO pins and prints to the segment display.
    \item Verilog module to send bytes of information over UART.
    \item Top-level verilog module that connects the previous modules for a standalone random number generator.
    \item Python module to read UART data and test for statistical randomness.
\end{enumerate}

\subsection{Random Algorithm Development Module}
Two algorithms will be investigated simultaneously throughout this project as a method of generating the random numbers. 

\subsubsection{The Middle-Square Algorithm}
The Middle-Square method, created by John Von Neumann, is an algorithm that takes an input number, then it squares the middle digits to get an output. That output is the next seed and the process continues in a linear feedback fashion.\cite{vonNeumann} 

\begin{figure}[H]
\centering
\includegraphics[scale=0.1]{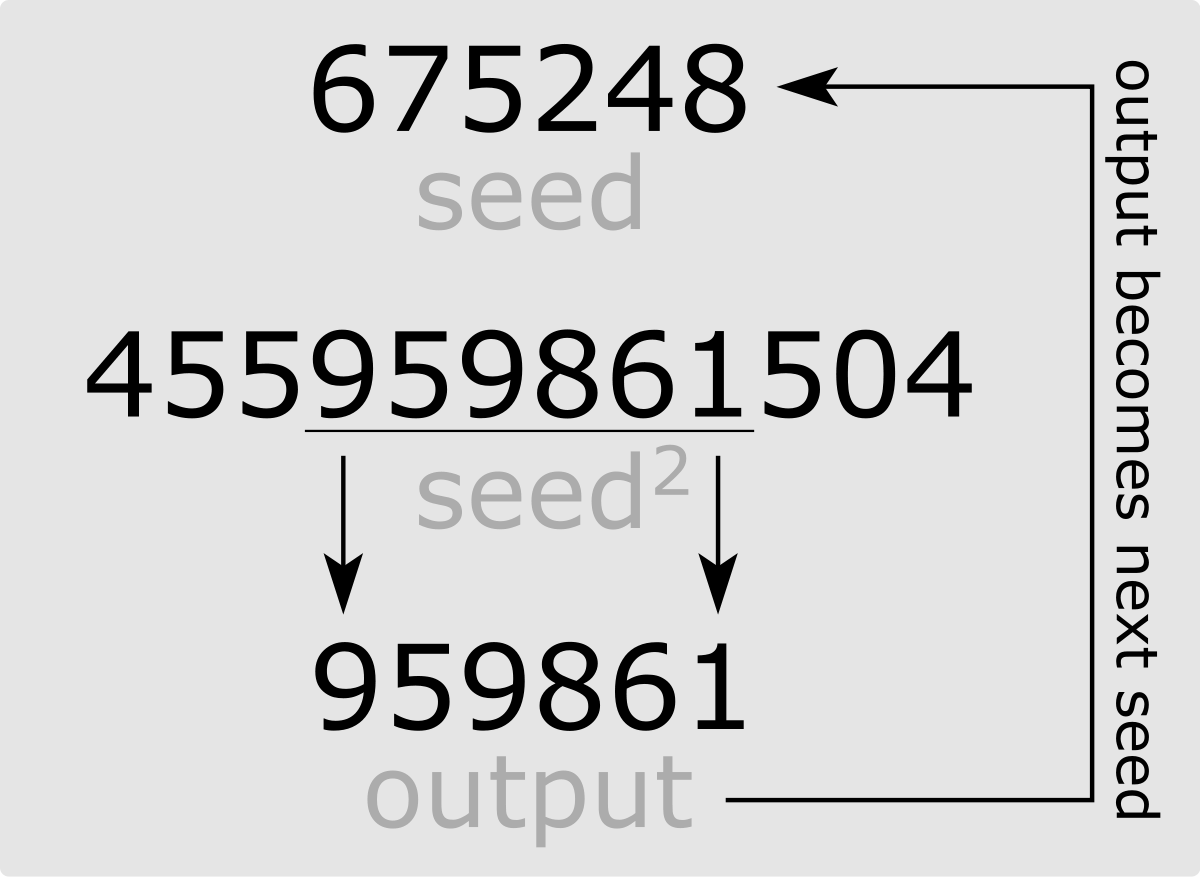}
\caption{Middle-Square Method}
\label{fig:ms_algo}
\end{figure}

Implementing the middle square method is very simple in verilog. Since this design aims to produce 32-bit random numbers, a seed value that is 32-bits long must be taken as an input. Then the middle digits can be squared and produce the output as the next random number. 
\begin{figure}[H]
\begin{lstlisting}[language=Verilog]
`timescale 1ns / 1ps
module middle_square (seed, rand);

    input [31:0]seed; //input seed is a 32 bit number
    output [31:0]rand; //output will be a 32 bit number
    
    wire [15:0] middle = seed[23:8]; //take the middle 16 digits of input
    wire [31:0] rand_out = middle * middle; //square the middle digits
    
    //assign output
    assign rand = rand_out; //output new 32 bit number
endmodule
\end{lstlisting}
\caption{Verilog for Middle-Square Method}
\end{figure}

\subsubsection{XORshift Algorithm}
The XORshift method, created by George Marsaglia, is an extremely fast and efficient algorithm for generating random numbers.  An XORshift algorithm works by taking the exclusive or (XOR) of a number with a shifted version of itself multiple times, before feeding back into the input again.\cite{xorshift} 

\begin{figure}[H]
\centering
\includegraphics[scale=0.20]{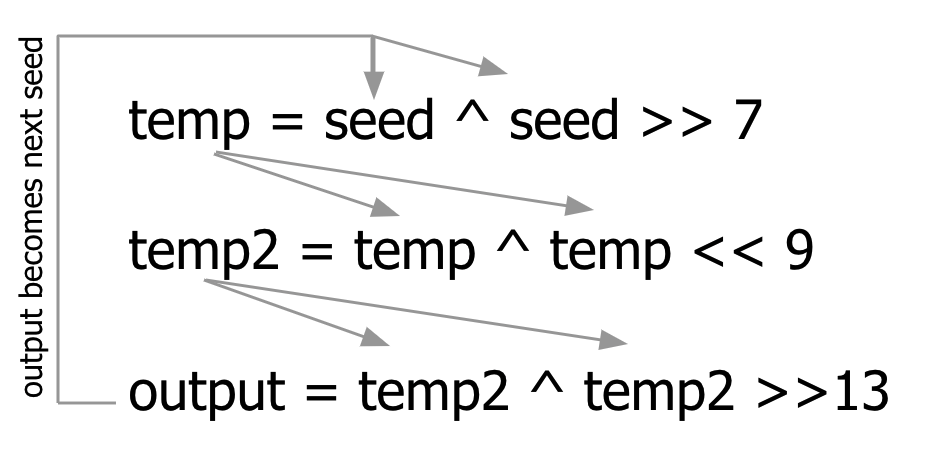}
\caption{XORshift Method}
\label{fig:xs_algo}
\end{figure}

Much like implementing the middle-square method, the XORshift requires a seed value that is 32-bits long. It then performs a XOR with a shifted version of itself several times before outputting the random number. 

\begin{figure}[H]
\begin{lstlisting}[language=Verilog]
`timescale 1ns / 1ps
module xorshift(seed, rand);

    input [31:0]seed; 
    output [31:0]rand; //output will be a 32 bit number
   
    //xorshift algorithm
    wire [31:0] temp = seed ^ seed >> 7;
    wire [31:0] temp2 = temp ^ temp << 9;
    wire [31:0] temp3 = temp2 ^ temp2 >>13;
    wire [31:0] rand_out = temp3;
    
    //assign output
    assign rand = rand_out; //output new 32 bit number
endmodule
\end{lstlisting}
\caption{Verilog for XORshift Method}
\end{figure}

\subsubsection{Testbench for Random Algorithm Modules}
The next step is to test that both the Middle-Square and XORshift verilog modules are operating as intended. A verilog testbench can be used to setup a simple register and intialize a static starting seed, and then connect the output of the XORshift or Middle-Square to the seed input for subsequent iterations. 

\begin{figure}[H]
\begin{lstlisting}[language=Verilog]
`timescale 1ns / 1ps
module random_tb;

    reg [31:0]seed; 
    wire [31:0]rand;
   
    //instantiation of algorithms, must select one at a time 
    xorshift xs (seed, rand); //middle_square ms (seed, rand);
    
    initial begin
    #0 seed = 32'h12345678; //seed is inputted manually for now
    forever
    #10 seed = rand;
    end
endmodule
\end{lstlisting}
\caption{Verilog for XORshift Method}
\end{figure}

By using a manual input seed, a waveform output can be generated to verify that the algorithms are performing correctly over time. Note that the outputted random number is becoming the next seed input to each algorithm respectively. 

\begin{figure}[H]
\centering
\includegraphics[scale=0.3]{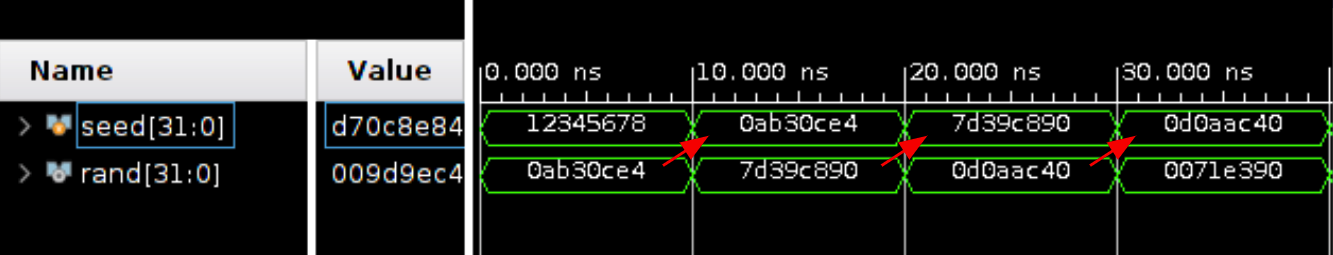}
\caption{Middle-Square Waveform}
\label{fig:ms_wav}
\end{figure}

\begin{figure}[H]
\centering
\includegraphics[scale=0.3]{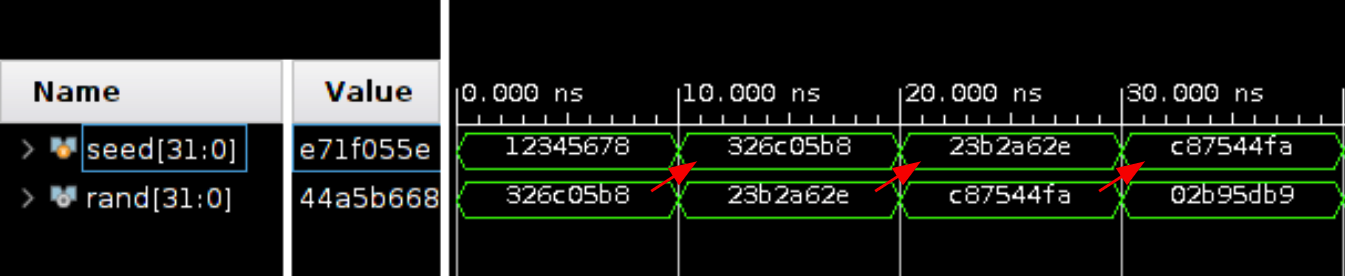}
\caption{XORshift Waveform}
\label{fig:xor_wav}
\end{figure}

\subsection{Nexys A7 GPIO Read Module}
The next module to will accomplish the task of reading a GPIO pin on the Nexys A7 development board for voltage and outputting the reading on the segment display. 

\subsubsection{Modify Board Constraints}
The Nexys A7 development board files obtained directly from Digilent contain a constraint file titled "Nexys-A7-100T-Master.xdc" which defines the mappings for physical pins into ports that are usable in the verilog design.\cite{Digilent} The digilent user guide has an in-depth walkthrough for how to setup the XADC, which is the on-board analog-to-digital converter linked to specific pins on the development board. 

\begin{figure}[H]
\begin{lstlisting}[language=Verilog]
#Pmod Header JXADC
set_property -dict {PACKAGE_PIN A14 IOSTANDARD LVCMOS33} [get_ports {vauxn3}];
set_property -dict {PACKAGE_PIN A13 IOSTANDARD LVCMOS33} [get_ports {vauxp3}]; 
set_property -dict {PACKAGE_PIN A16 IOSTANDARD LVCMOS33} [get_ports {vauxn10}]; 
set_property -dict {PACKAGE_PIN A15 IOSTANDARD LVCMOS33} [get_ports {vauxp10}]; 
set_property -dict {PACKAGE_PIN B17 IOSTANDARD LVCMOS33} [get_ports {vauxn2}]; 
set_property -dict {PACKAGE_PIN B16 IOSTANDARD LVCMOS33} [get_ports {vauxp2}]; 
set_property -dict {PACKAGE_PIN A18 IOSTANDARD LVCMOS33} [get_ports {vauxn11}]; 
set_property -dict {PACKAGE_PIN B18 IOSTANDARD LVCMOS33} [get_ports {vauxp11}]; 
\end{lstlisting}
\caption{Enable JXADC Pins in Board Constraints File}
\end{figure}

\subsubsection{Instantiating the XADC and Segment Display}
The XADC on the Nexys A7 is compatible with the Xilinx IP instantiation called XADC Wizard which is an easy to use wizard to configure the on-chip ADC of the FPGA.\cite{Xilinx} It allows us to connect physical pins on the board as wires of data into the design. The "do\_out" signal is a 15-bit bus, however this particular ADC only provides 12-bits of precison to read each differential input pair in range of $\pm 1.0 V$. These 12-bits will be converted to decimal values which will then be displayed on the segment display. The code for the seven segment display was used directly from Digilent's github repository for the NexysA7 development example projects.\cite{Digilent_git}

\begin{figure}[H]
\begin{lstlisting}[language=Verilog]
 always @ (posedge(CLK100MHZ))
      begin
        //refresh segment display every 1/10th second  
        if(count == 10000000)begin
        decimal = data >> 4;
        begin
            decimal = decimal * 250000;
            decimal = decimal >> 10;
            dig0 = decimal % 10;
            decimal = decimal / 10;
            dig1 = decimal % 10;
            decimal = decimal / 10;
            dig2 = decimal % 10;
            decimal = decimal / 10;
            dig3 = decimal % 10;
            decimal = decimal / 10;
            dig4 = decimal % 10;
            decimal = decimal / 10;
            dig5 = decimal % 10;
            decimal = decimal / 10;
            dig6 = decimal % 10;
            decimal = decimal / 10;
            count = 0;
            end
        count = count + 1;//increment segment display counter 
        Address_in <= 8'h13; //channel address corresponding to NexysA7 XADC pins AD3P/N
        end
    end

//xadc instantiation
   xadc_wiz_0  XLXI_7 (.daddr_in(Address_in),
                     .dclk_in(CLK100MHZ), 
                     .den_in(enable), 
                     .di_in(0), 
                     .dwe_in(0), 
                     .busy_out(),                    
                     .vauxp2(vauxp2),
                     .vauxn2(vauxn2),
                     .vauxp3(vauxp3),
                     .vauxn3(vauxn3),
                     .vauxp10(vauxp10),
                     .vauxn10(vauxn10),
                     .vauxp11(vauxp11),
                     .vauxn11(vauxn11),
                     .vn_in(vn_in), 
                     .vp_in(vp_in), 
                     .alarm_out(), 
                     .do_out(data), 
                     .reset_in(0),
                     .eoc_out(enable),
                     .channel_out(),
                     .drdy_out(ready));
                     
      //segment display instantiation
      DigitToSeg segment1(.in1(dig0),
                         .in2(dig1),
                         .in3(dig2),
                         .in4(dig3),
                         .in5(dig4),
                         .in6(dig5),
                         .in7(dig6),
                         .in8(dig7),
                         .mclk(CLK100MHZ),
                         .an(an),
                         .seg(seg));  
                         
endmodule
\end{lstlisting}
\caption{Verilog for GPIO Module}
\end{figure}

\subsubsection{Testing GPIO Module}
The next step is to test that both the Nexys A7 board and verilog modules are correctly measuring voltage and displaying onto the segment display. To accomplish this, a simple voltage divider can be setup to use the 3.3V supply directly from the Nexys A7 board and divide the voltage to something in the range (below 1.0V). Following the formula for a two-resistor voltage divider, the following circuit is developed. 

\begin{equation}
    V_{out} = ( V_{in}\cdot R_1) / (R_2 + R_1)
\end{equation}
\begin{equation}
    V_{out} = ( 3.3V\cdot 10k) / (1k + 10k) = 0.3V
\end{equation}

\begin{figure}[H]
\centering
\includegraphics[scale=0.25]{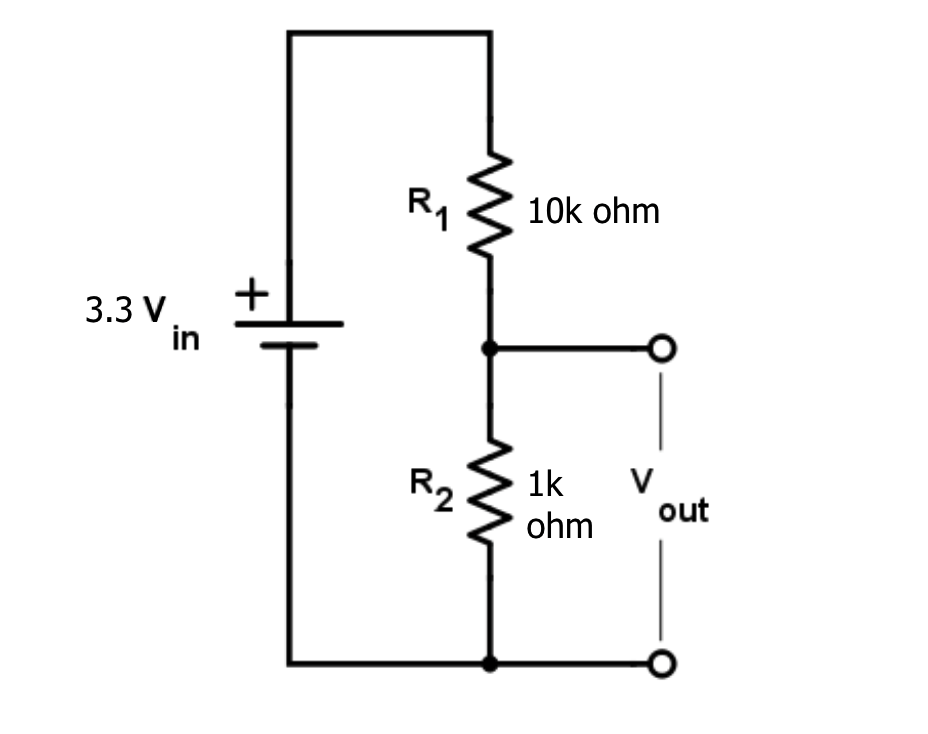}
\caption{Voltage Divider Circuit}
\label{fig:volt_div}
\end{figure}

\begin{figure}[H]
\centering
\includegraphics[scale=0.2]{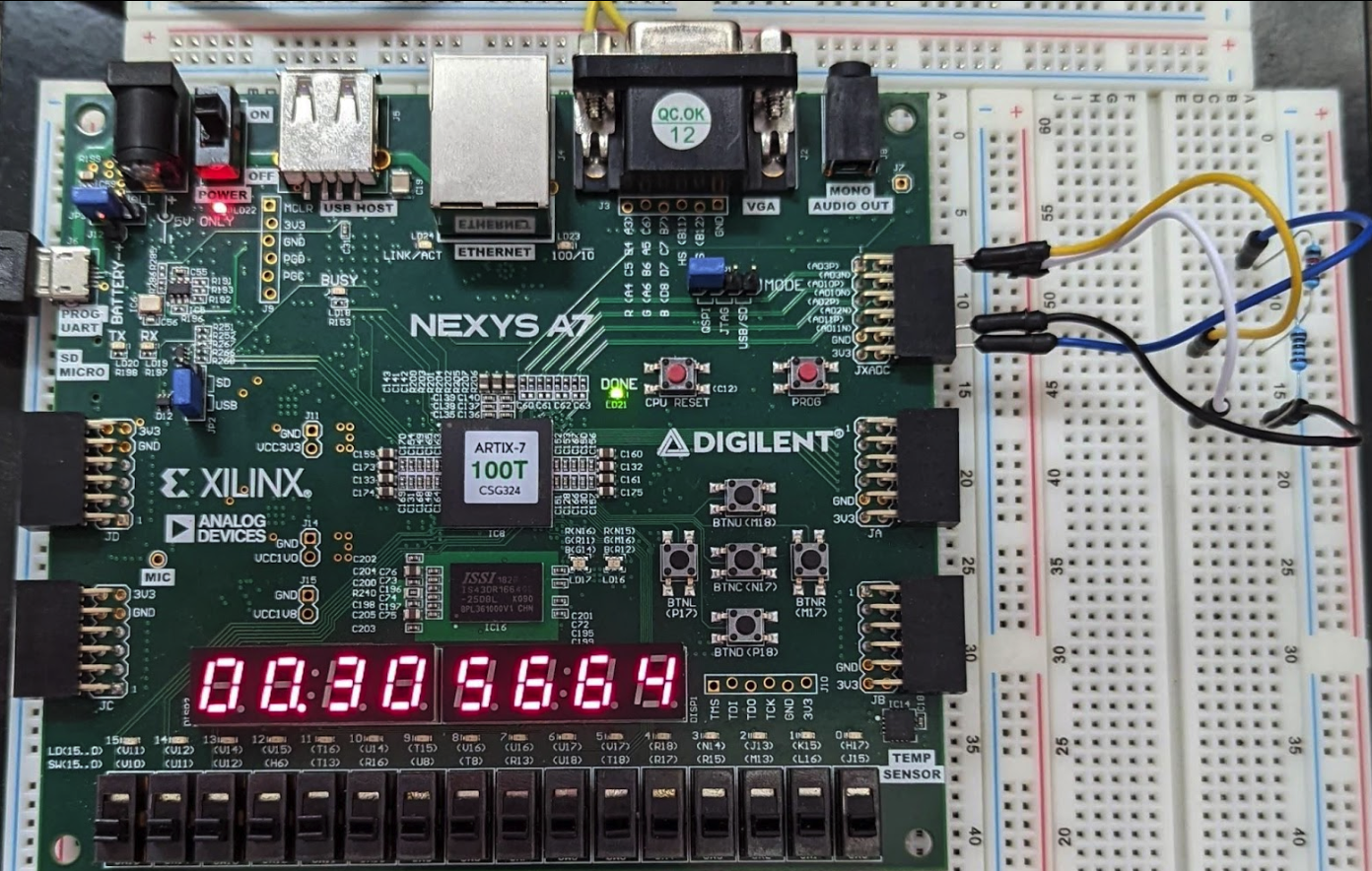}
\caption{GPIO Module Test on Nexys A7}
\label{fig:nex_div}
\end{figure}

The results on the segment display indicate a reading of 0.3V which means that the verilog module successfully utilizes analog voltage inputs and converts them to digital data reliably. This will be important for measuring electrical noise as the input seed to the final random number generator. 

\subsection{UART Module}
UART (universal asynchronous receiver-transmitter) is a communication protocol between devices that operates by sending bytes of information.\cite{uart} UART can be configured to work with serial protocols, which in the case of the Nexys A7 is Serial RS232 to USB transmitter/receiver via a dedicated FTDI\texttrademark  chip.\cite{Digilent}

\subsubsection{Modify Board Constraints}
The board constraints file "Nexys-A7-100T-Master.xdc" must be modified again to enable the pins to the UART interface. 

\begin{figure}[H]
\begin{lstlisting}[language=Verilog]
#USB-RS232 Interface
set_property -dict {PACKAGE_PIN C4 IOSTANDARD LVCMOS33} [get_ports {UART_TXD_IN}];
set_property -dict {PACKAGE_PIN D4 IOSTANDARD LVCMOS33} [get_ports {UART_RXD_OUT}]; 
\end{lstlisting}
\caption{Enable UART/RS232 Interface in Board Constraints File}
\end{figure}

\subsubsection{UART TX CTRL Module}
Digilent's github repository also provides a UART transmit VHDL (virtual hardware description language) code to use with the Nexys A7 Board.\cite{Digilent_git} VHLD is similar to verilog and can be used within the same design as long as the signals and connected nets are correctly configured. 

\begin{figure}[H]
\begin{lstlisting}[language=VHDL]
entity UART_TX_CTRL is
    Port ( SEND  : in  STD_LOGIC;
           DATA  : in  STD_LOGIC_VECTOR (7 downto 0);
           CLK   : in  STD_LOGIC;
           READY : out STD_LOGIC;
           UART_TX : out  STD_LOGIC);
end UART_TX_CTRL;
\end{lstlisting}
\caption{UART Transmit Control VHDL}
\end{figure}

To connect this module to a verilog design the following signals must be accounted for:
\begin{itemize}
    \item \textbf{SEND}: Triggers a send operation, set signal high for a single clock cycle to trigger a send. DATA must be valid when SEND is asserted. SEND should not be asserted unless READY is high.
    \item \textbf{DATA}: 8 bits (1 byte). The UART code will add a stop bit to the final message. 
    \item \textbf{CLK}: 100 MHz system clock
    \item \textbf{READY}: This signal goes low once a send operation starts, remains low until complete and ready to send next byte
    \item \textbf{UART\_TX}: Route this to the TX pin on the Nexys A7 (constraints .xcd file)
\end{itemize}

\subsubsection{Verilog Logic to Send 4 Bytes}
Since there are several signals that have to be asserted according to the UART transmit code, it means that there must be logic in the top-level verilog code to account for asserting signals and waiting for the READY signal before sending a byte of information. Since the goal is to generate 32-bit random numbers, this means that each number will need to be divided into 4 bytes. UART can only transmit one byte at a time, so each byte will need to be passed sequentially to get the entire random number transmitted. 
\begin{figure}[H]
\begin{lstlisting}[language=verilog]
   //UART controls
   wire uart_ready;
   reg send_uart;
   reg [7:0] byte1_to_send;
   reg [7:0] byte2_to_send;
   reg [7:0] byte3_to_send;
   reg [7:0] byte4_to_send;
   reg [7:0] uart_byte = 32'h12345678;
   reg [31:0] byte_count;  
   
    //uart instantiation
      UART_TX_CTRL uart( .SEND(send_uart),
                         .DATA(uart_byte),
                         .CLK(CLK100MHZ),
                         .READY(uart_ready),
                         .UART_TX(UART_RXD_OUT)); 
                         
 //send random number byte over UART
       //grab latest rand, store in temp regs until uart is done transmitting 
       //all 4bytes (takes 868 clock cycles per byte @115200 baud)
        byte1_to_send = rand_out[7:0];
        byte2_to_send = rand_out[15:8];
        byte3_to_send = rand_out[23:16];
        byte4_to_send = rand_out[31:24];
            
        //choose first byte to send
        if(byte_count == 0)begin
            uart_byte = byte1_to_send;
            end 
         //chose next byte to send
        if(byte_count == 1)begin
            uart_byte = byte2_to_send;
            end
         //chose next byte to send   
         if(byte_count == 2)begin
            uart_byte = byte3_to_send;
            end  
         //chose next byte to send   
         if(byte_count == 3)begin
            uart_byte = byte4_to_send;
            byte_count = 0;
            end  
            
       //Check if UART ready signal. If ready, load byte values 
       //and assert send signal for each byte 
        if(uart_ready == 1)begin 
            send_uart = 1'b1;    
            byte_count = byte_count + 1;
            end 
        //if UART not ready, do not send next byte
        else begin
            send_uart = 1'b0;
        end
      end                         
\end{lstlisting}
\caption{UART Transmit 32-bits}
\end{figure}

\subsubsection{Testing UART Module}
The next step is to test that both the Nexys A7 board and verilog modules can correctly transmit a 32-bit number over the USB/RS232/UART interface. This can be setup by using a USB cable connected from the Nexys A7 board to a computer's USB port. To set up serial communication, Python and the pySerial library make this process quick and easy.\cite{pyserial}

\begin{figure}[H]
\begin{lstlisting}[language=Python]
import serial

#Open serial port
ser = serial.Serial(
    port='/dev/tty.usbserial-210292ABF4F61', #The FTDI serial/USB device
    baudrate=115200, #baud rate that my UART_TX is using
    parity=serial.PARITY_NONE, #no parity used 
    stopbits=serial.STOPBITS_ONE, #single stop bit to signal end of data
    bytesize=serial.EIGHTBITS, #byte size is eight bits
    timeout = None
)
hex_array = [] #create a python list/array to store all of the bytes in

for i in range(0,490000): #will store a total of 490,000 32-bit numbers
    hex_array.append(ser.read(4).hex()) #concatenates 4bytes per list entry in hex
    
ser.close() #close serial port
#write the data to a text file
a_file = open("rand.txt", "w")
for row in hex_array:
    a_file.write("".join(row) +"\n")
a_file.close()
\end{lstlisting}
\caption{Serial Data Receive - Python}
\end{figure}

Now, the Nexys A7 board can be programmed with the UART module verilog code and connected to the computer running Python. The python program will begin the serial data communication and store the data into a text file. 
\begin{figure}[H]
\begin{lstlisting}[language=Python]
0x12345678
0x12345678
0x12345678
.
.
. #text file continues up to 490,000 lines
\end{lstlisting}
\caption{Output Text File}
\end{figure}

The result of this test is the "uart\_byte" which was set to "0x12345678" repeated 490,000 times and a text file ~4.4MB in size. This means the FPGA can successfully transmit a 32-bit number continuously over UART to python which will be important later for analyzing the generated random numbers. 

\subsection{Top-level Verilog Design}
Since there are now have three working modules---the random algorithm module, the GPIO module, and the UART module, the next step is to combine them into a single top level verilog design to generate random numbers. To accomplish this the physical jumper cables should be removed from the NexysA7 XADC pin headers so that the differential input pairs are reading ambient electrical noise. This is the event that is truly random that will be measured to use as the input seed into the random number algorithms. Once that is connected, the random number can be sent to the segment display as well as the UART transmit code to send each random number generated over serial for further analysis. 

\begin{figure}[H]
\begin{lstlisting}[language=Verilog]
      always @ (posedge(CLK100MHZ))
      begin
        //get initial seed from GPIO/XADC pin measurement to start
        if(seed_count == 0)begin
            seed_in = data*data;
            end
        else begin
            seed_in = rand_out;
            end
       //set new seed to be the output of last generated random number (linear feedback!)
        if(seed_count < 1000)begin
            seed_in = rand_out;
            end
       //else refresh seed from GPIO/XADC pin measurement every 1000 clock cycles 
        else begin
            seed_in = data << 8;
            seed_count = 0;
            end
        seed_count = seed_count +1;
        //refresh segment display every 1/10th second  
        if(count == 10000000)begin
            dig0 = rand_out[3:0];
            dig1 = rand_out[7:4];
            dig2 = rand_out[11:8];
            dig3 = rand_out[15:12];
            dig4 = rand_out[19:16];
            dig5 = rand_out[23:20];
            dig6 = rand_out[27:24];
            dig7 = rand_out[31:28];
            count = 0;
            end
        count = count + 1;//increment segment display counter 
        Address_in <= 8'h13; //channel address corresponding to NexysA7 XADC pins AD3P/N
       //send random number byte over UART grab latest rand, store in temp regs until
       //uart done transmitting 4bytes (takes 868 clock cycles per byte @115200 baud)
        byte1_to_send = rand_out[7:0];
        byte2_to_send = rand_out[15:8];
        byte3_to_send = rand_out[23:16];
        byte4_to_send = rand_out[31:24];
        //choose first byte to send
        if(byte_count == 0)begin
            uart_byte = byte1_to_send;
            end 
         //chose next byte to send
        if(byte_count == 1)begin
            uart_byte = byte2_to_send;
            end
         //chose next byte to send   
         if(byte_count == 2)begin
            uart_byte = byte3_to_send;
            end  
         //chose next byte to send   
         if(byte_count == 3)begin
            uart_byte = byte4_to_send;
            byte_count = 0; //reset byte count back to zero
            end  
       //Check if UART ready signal 
       //if ready, load byte values 
       //and assert send signal for each byte 
        if(uart_ready == 1)begin 
            send_uart = 1'b1;    
            byte_count = byte_count + 1; //increment to next byte
            end 
        //if UART not ready, do not send next byte
        else begin
            send_uart = 1'b0;
        end
      end
\end{lstlisting}
\caption{Top-Level Verilog Design - Nexys A7 Random Number Generator}
\end{figure}

The verilog module hierarchy is as shown:
\begin{figure}[H]
\centering
\includegraphics[scale=0.35]{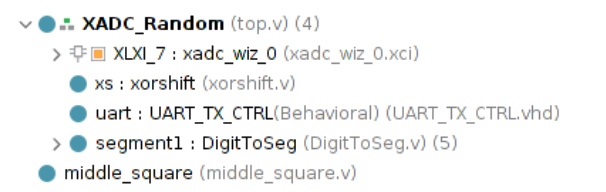}
\caption{Verilog Top-Level Module Hierarchy}
\label{fig:hierarchy}
\end{figure} 

\subsubsection{Testing Top-Level Verilog Design}
The next step is to test that the top-level design is behaving as anticipated. It should now be generating a new 32-bit random number every clock cycle, printing numbers to the segment display, and transmitting the numbers over the UART interface to python. Since the GPIO module is providing electrical noise as the input seed to the random algorithms, they will continue to generate random numbers continuously and be refreshed with a new seed every 1000 clock cycles. 

\begin{figure}[H]
\centering
\includegraphics[scale=0.25]{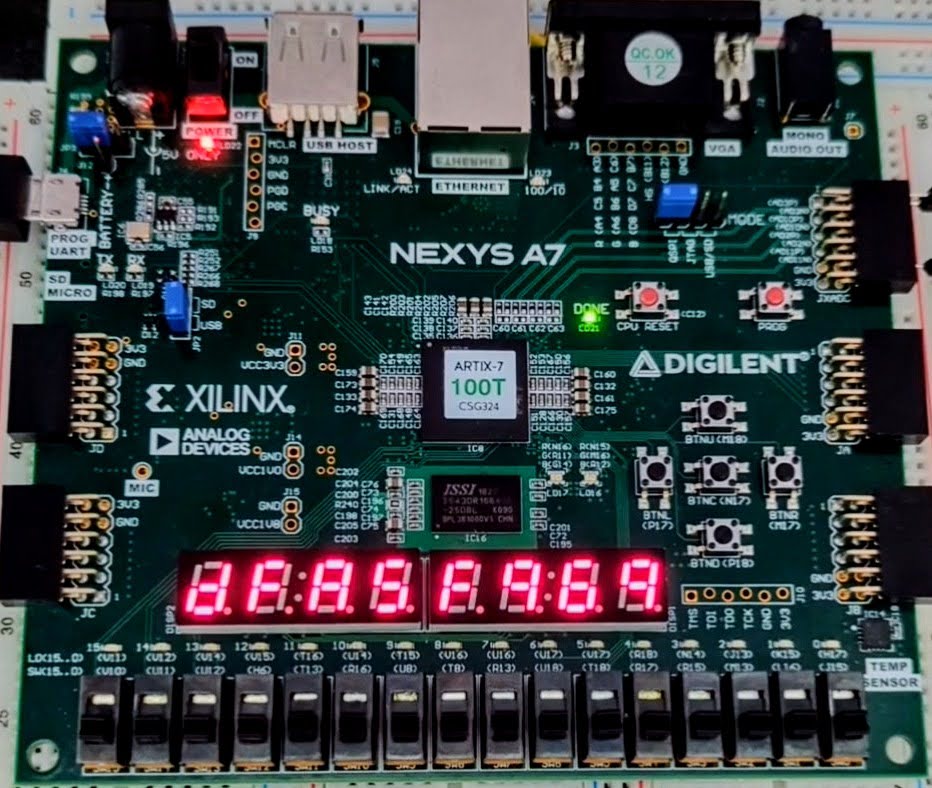}
\caption{Nexys A7 Display Output - 32bit Numbers Displayed in Hex}
\label{fig:nexys_seg}
\end{figure} 

The python program will begin the serial data communication and store the data into a text file. 
\begin{figure}[H]
\begin{lstlisting}[language=Python]
0x9d9696ae
0xb87be462
0x5a5ca400
0x8080c22d
0xbaf8783d
0x2488f3c2
.
.
. #text file continues up to 490,000 lines
\end{lstlisting}
\caption{Output Text File - Random Numbers}
\end{figure}

The result of this test is a text file filled with 490,000 randomly generated numbers represented in hexadecimal format. This means the system can successfully transmit a 32-bit random numbers over UART to python which can now be analyzed for statistical randomness. 

\subsection{Python Module}
The final module is a python program to analyze the statistical randomness of the generated random numbers. The following program opens the text file filled with randomly generated numbers, then it iterates over the list of numbers and normalizes each number to be less than 1.0 converting each one to a float. If the set of random numbers is truly random, then the distribution of those numbers between 0 and 1 should be uniform. To demonstrate this in a clear way, each number will be evaluated. If the number is less than 0.5, then a single pixel on a blank canvas will be colored black. If the number is greater than 0.5, then a single pixel on the same canvas will be colored white. This process repeats for all 490,000 numbers until a 700 by 700 pixel image is created to represent the statistical randomness of the numbers.  

\begin{figure}[H]
\begin{lstlisting}[language=Python]
# Libraries
from tkinter import *
# Global variables
y = 0  # used for y axis in tkinter visual model
x = 0  # used for x axis in tkinter visual model
# Create a new graphic window in tkinter
window = Tk()
graphic = Canvas(window, width=700, height=700, background='white')

# Gather random numbers from text file, store into array
def get_rands():
    num_array = []
    h = open('rand.txt', 'r')
    content = h.readlines()
    
    for line in content:
        temp = int(line, 16) #convert hex string to integer
        num_array.append(temp)
    return num_array

# Convert random number to float less than 1
def rand_float(rand):
    # divide by the maximum 32 bit hex number number to turn current random number into type float
    float_rand = (rand) / int('ffffffff', 16)
    return (float_rand)

# Main Program
#create float array of random numbers
results = get_rands()

for num in range((len(results))):
    results[num] = rand_float(results[num])
i=0
# Generate psuedo random numbers and convert to on/off pixels in tkinter visual model (490,000 RNs)
while(y < 700):  # row index
    for x in range (700): #column index
        # check if number is even, should be a 50% chance
        if(results[i] < 0.5):
            # if true, generate black pixel
            graphic.create_rectangle(x, y, x+1, y+1, outline="", fill="black")
            i = i+1
        else:
            # else, generate white pixel
            graphic.create_rectangle(x, y, x+1, y+1, outline="", fill="white")
            i = i+1
    graphic.pack()  # compile tkinter graphics
    y = y + 1  # increment to the next row.
window.mainloop()  # write graphics to tkinter graphics window and display
\end{lstlisting}
\caption{Statistical and Graphical Random Test in Python}
\end{figure}

\subsection{Testing the Python Statistical Randomness Program}
To test this final module, we can feed in a list of numbers that are generated with a pseudo-random number generator first. We can use the middle-square method from before, but instead of feeding in a seed from our XADC reading, we will set it to something static like 0x19238433. Recall that the middle-method will take this seed and square the middle digits making that the output, and the next iteration's seed.

\begin{figure}[H]
\centering
\begin{subfigure}{.5\textwidth}
  \centering
  \includegraphics[width=.5\linewidth]{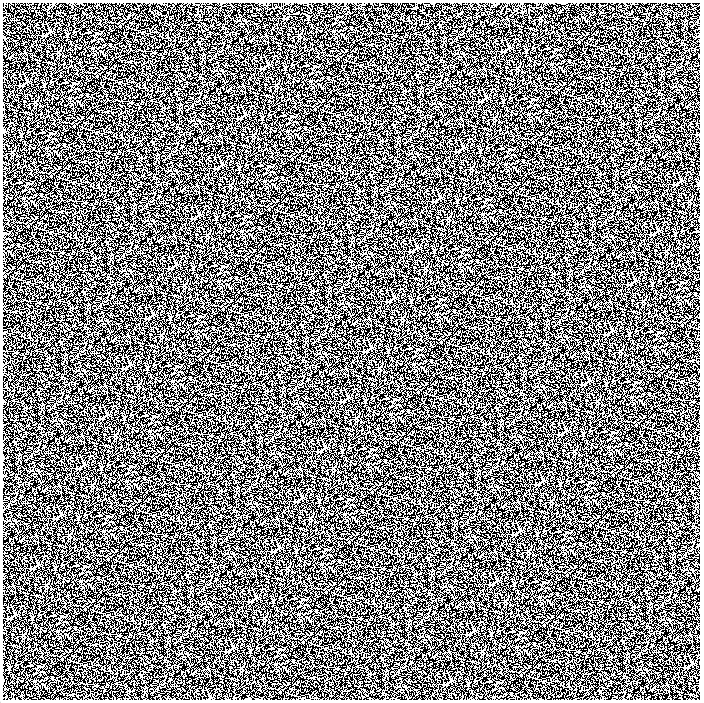}
  \caption{Pseudo-Random Output - Seed=0x19238433}
  \label{fig:sub1}
\end{subfigure}%
\begin{subfigure}{.5\textwidth}
  \centering
  \includegraphics[width=.5\linewidth]{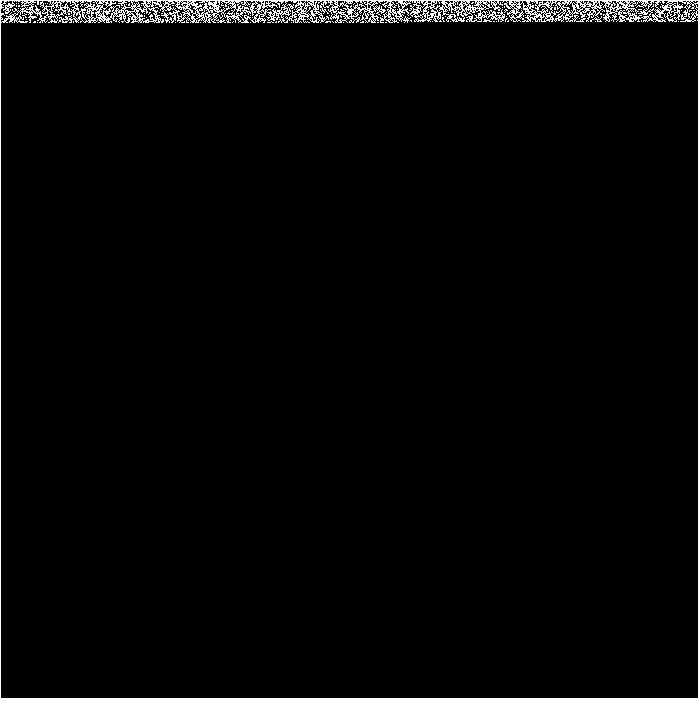}
  \caption{Pseudo-Random Output - Seed=0x20118433}
  \label{fig:sub2}
\end{subfigure}
\caption{Python Statistical Randomness Test Graphical Output}
\label{fig:pseudo_stats}
\end{figure}

Some observations to note about this output right now is that it looks something like television static. What is being shown is the overal distribution of numbers greater than or less than 0.5 respectively where each pixel colored is one case or the other. While the result in \textbf{Figure \ref{fig:sub1}} may appear random at first glance, if you look (or move your eyes a bit further from this image), you might see there are several diagonal lines or patterns that start to appear. What this means at a high level is that the distribution of random numbers is actually \textit{repeated} after some amount of time. That is why you can see a pattern in the distribution after some amount of time. 

\textbf{Figure \ref{fig:sub2}} shows a unique case where the result is a feedback loop. Recall that the middle-square method takes the middle digits and squares them for the next input. What happens in the case where the middle digits are all zero? Since $0^2 = 0$ the algorithm ends up in a loop that will produce the number 0 \textit{forever}. Since both of these tests were Pseudo-random numbers, it is expected that they cannot produce truly random numbers forever, so this result is to be expected. 

\section{Results}
The final step of this project is to analyze the data from the random numbers generated by the Nexys A7 board using the GPIO measurement of ambient electrical noise as the input see to the random number algorithms. 

\begin{figure}[H]
\centering
\begin{subfigure}{.5\textwidth}
  \centering
  \includegraphics[width=.5\linewidth]{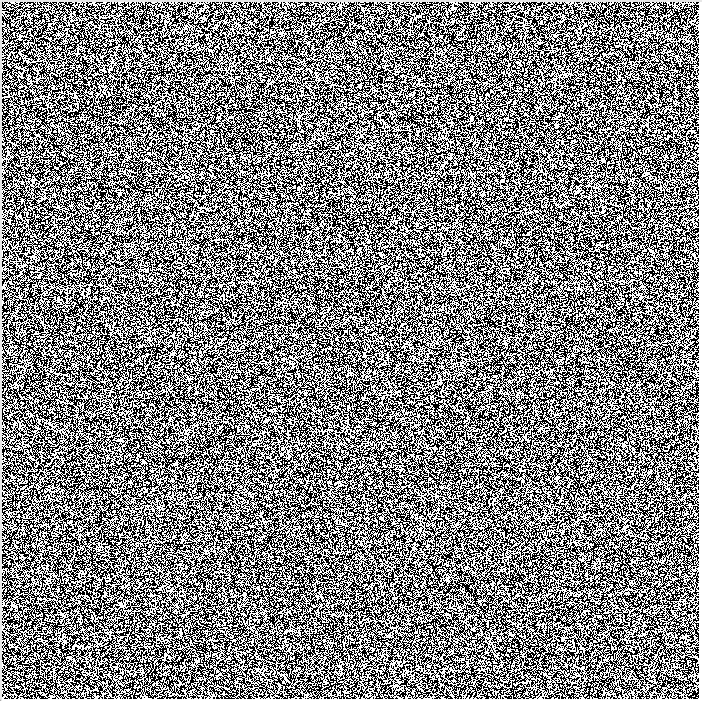}
  \caption{XORshift Random Output}
  \label{fig:subxor}
\end{subfigure}%
\begin{subfigure}{.5\textwidth}
  \centering
  \includegraphics[width=.5\linewidth]{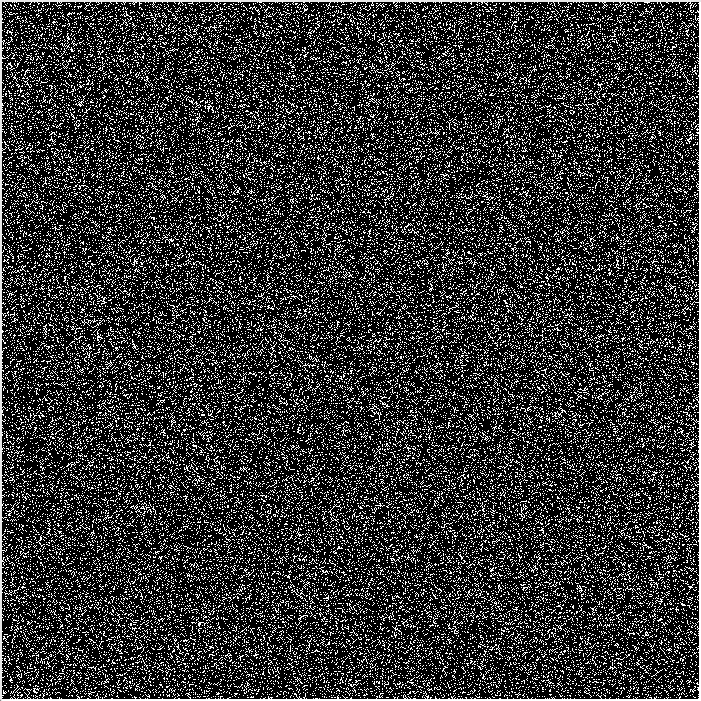}
  \caption{Middle-Square Random Output}
  \label{fig:subms}
\end{subfigure}
\caption{Python Statistical Randomness Test Graphical Output}
\label{fig:pseudo_stats_real}
\end{figure}

Some observations to note about the final outcome. \textbf{Figure \ref{fig:subxor}} shows a picture resembling random television static, only this time there are no discernible lines or patterns in the output when compared to \textbf{Figure \ref{fig:sub1}}. This is great news because it means that at a high-level it can be said with confidence that this method does not generate random numbers in a repeated pattern in any statistically significant way. The overall color of the image is gray meaning that it is likely that 50\% of the numbers are greater than 0.5 and the remaining 50\% are less than 0.5. This supports the idea that the numbers are truly random since there are no patterns and the distribution is uniform. 

Observing \textbf{Figure \ref{fig:subms}} one can see that it is much darker than the XORshift outcome. What this means is that this method tends to generate numbers that are overall smaller over time. While there is no discernible pattern, the fact that the distribution of these random numbers tends to be biased towards smaller numbers makes this method not truly random for generating all numbers possible with equal probability. This might be due to the fact that there are many 32-bit numbers where the middle 16bits can equal zero. In-fact, if this is the case at any point during the loop, the middle-square method will enter that feedback cycle of generating zero until a new seed is refreshed every 1000 clock cycles. This is likely the case.

Overall if one is looking for true random numbers, the most promising method looks to be the XORshift method presented. 

\subsection{Design Limitations}
One limitation of this design and project is related to UART transmission. The design itself generates a new random number each clock cycle, however UART is only able to transmit the 4-bytes for each number at a limited baud rate of 115200 symbols per second. At a clock rate of 100MHz it turns out that the Nexys board can only transmit an entire 32-bit number over the UART interface once every ~3500 clock cycles. This ultimately means that the analyzed random numbers are only getting a sample of the generated numbers once every 3500 generated numbers. Ideally, it would be best to analyze every single number generated by ensuring we either store the data or can transmit synchronously with the clock. However, the graphical outputs produced as a result of this project can still be valid since a statistical sample of a random set should still remain random. 

\section{Conclusions}
This project proved several things:
\begin{itemize}
    \item Randomizing algorithms can be implemented directly in hardware using simple logical gates
    \item Linear feedback-type machines are possible to create continuous operations where the output of one function is the next iteration's input
    \item XADC I/O support was added to the Nexys A7 board with the ability to read voltages accurately
    \item A functioning UART interface over USB/RS232 serial from the Nexys A7 can be used to transmit data from the FPGA directly to another computer
    \item Randomness of the generated numbers can be verified by graphically plotting the distribution
    \item The XORshift method seems to be superior to the Middle-square method for generating random numbers
\end{itemize}

\section{Future Recommendations}
As mentioned in \textbf{Section 4.1}, UART transmission proved to be a bottle neck in retrieving and analyzing all random numbers generated by the Nexys A7 board. Some recommendations for improving this project or developing further could be:
\begin{enumerate}
    \item Instead of UART transmission in real-time, one might store the random numbers in a memory location on the FPGA first. Then use UART to iterate over each memory location an send all bytes of information without losing any to delays associated with UART speed.
    \item This project is setup to be instantiated into other verilog modules. Since it produces 32-bit random numbers each clock cycle perhaps one could implement this in MIPS and use the random number to do tasks like "store word", "load word", use the random numbers in arithemtic instructions, or perhaps to create a routine to fetch addresses randomly. 
\end{enumerate}

\bibliographystyle{IEEEtran}
\bibliography{refs}

\end{document}